# Delivering SKA Science

**P.J.Quinn[1,2], T. Axelrod[3], I. Bird[4], R. Dodson[2], A. Szalay[5], A. Wicenec[2]**
[2]*International Centre for Radio Astronomy Research (ICRAR),* [3]*LSST Project,* [4]*LHC Computing Grid Project, CERN,* [5]*Institute for Data Intensive Science, Johns Hopkins University*
*E-mail:* peter.quinn@icrar.org

The SKA will be capable of producing a stream of science data products that are Exa-scale in terms of their storage and processing requirements. This Google-scale enterprise is attracting considerable international interest and excitement from within the industrial and academic communities. In this chapter we examine the data flow, storage and processing requirements of a number of key SKA survey science projects to be executed on the baseline SKA1 configuration. Based on a set of conservative assumptions about trends for HPC and storage costs, and the data flow process within the SKA Observatory, it is apparent that survey projects of the scale proposed will potentially drive construction and operations costs beyond the current anticipated SKA1 budget. This implies a sharing of the resources and costs to deliver SKA science between the community and what is contained within the SKA Observatory. A similar situation was apparent to the designers of the LHC more than 10 years ago. We propose that it is time for the SKA project and community to consider the effort and process needed to design and implement a distributed SKA science data system that leans on the lessons of other projects and looks to recent developments in Cloud technologies to ensure an affordable, effective and global achievement of SKA science goals.

---

[1]  Speaker

# 1. Introduction

Over the past 20 years, the concept of what constitutes an observatory has grown to combine three central elements – the collectors, the detectors and the data. The delivery of data to astronomers in a form and variety that enables and accelerates the research process has become part of the fabric of new facilities in an attempt to achieve a maximal scientific return on an ever increase commitment of public funds. The capture and delivery of data of the right type, with the right quality, at the right time and with the right volume is a complex dance between collectors, detectors, those responsible for a facility and astronomers responsible for doing and communicating science. This complex network of relationships and resources needs to be identified, costed, prioritized and addressed as part of designing and building the SKA facility. The data and processing systems that enable SKA science (collectively called SKA-Data) should be regarded at the same level of scientific importance, visibility and criticality as the collectors/detectors known as SKA-LOW, SKA-MID and SKA-SUR. The ultimate scientific impact of SKA will be determined by the way in which we assign priorities and resources to all four of these elements of the SKA, recognizing that SKA-data supports all three collecting/detecting systems.

SKA is being showcased internationally as an Exa-scale computing project and considerable excitement and expectation is being generated in the scientific and industrial community by the associated technological and discovery opportunities. Today we know that Exa-scale enterprises require Google-like resources and we need to understand how astronomy will scale to, and afford, challenges of this magnitude. In this chapter we will examine the SKA-data requirements as determined by the set of SKA Key Science Projects. The existing SKA Phase 1 baseline design coupled with HI survey requirements already implies data volumes of processed and accumulated products for a single survey to be in excess of 1 ExB. We will discuss the projected costs and challenges for the production and delivery of SKA science products. This analysis will draw upon the experience gained across a number of major research efforts (Sloan Digital Sky Survey, Large Hadron Collider and the Large Synoptic Survey Telescope). We will draw attention to choices that may need to be made on the scope and capabilities of SKA-data and what those choices may imply for funding sources and operational models. Finally we will outline technological trends and innovations that may create new opportunities to do science with the SKA and which could change the design approach we take to SKA-data within this decade.

# 2. SKA Data Requirement

We begin our assessment of the scale of SKA's data requirements with an analysis of the data production for SKA1 based on a set of Key Science Projects (KSP, Lazio, 2011). For the 3 component arrays, SKA-LOW, SKA-MID and SKA-SUR, we take the following global parameters from the baseline design plus the addendum:

|  | SKA-LOW | SKA-MID | SKA-SUR |
|---|---|---|---|
| Nant | 1024 | 254 | 96 |
| Bmax (km) | 100 | 200 | 50 |
| Dstation (m) | 35 | 15 | 15 |
| BWmax(MHz) | 250 | 1024 | 500 |
| Nbeams | 1 | 1 | 36 |

*Table 1: Assumed baseline design parameters*

For each KSP, the fiducial frequency is calculated based on the maximum and minimum frequency requested. This calculation produces figures that are reasonably close to the fiducial frequencies used in the baseline design. The figures are then used to calculate the sampling times and channel widths for the array to prevent temporal or bandwidth smearing. Oversampling factors for the time, frequency and images are $\varepsilon_\tau$ 11.2, $\varepsilon_\nu$ 8 and $\varepsilon_{image}$ 4, respectively. The bytes per sample for the raw data, the visibilities and images are $B_{raw}$ 1, $B_{vis}$ 8 and $B_{image}$ 4, respectively. Dual polarisations are recorded in all cases. The antennas are assumed to be distributed following the configurations proposed in the baseline design. This configuration affects the possible gains from the differential averaging based on baseline lengths. Based on these inputs, we have settled on averaging factors of 14.7, 13.8 and 4.81 for SKA-LOW, SKA-MID and SKA-SUR respectively. We assume in all cases that 6 hours of data is used for each image and that projects all receive 1 kHr for the SKA-LOW observations and 10 kHr for the SKA-MID and SKA-SUR observations in order to produce comparable results. Where only postage-stamps are required we have retained 100 pixels (or 25 resolution samples) around the target in the two on-sky dimensions and 1000 channels in velocity. Where only the continuum data is required the number of channels is reduced to 1000, to allow recovery of Rotation Measures. In the baseline design there is a limit of $2^{18}$ for the number of channels that can be formed. Therefore for some of the cases with narrow velocity channel width requirements, we have limited the correlated bandwidth to that which can be supported. This affects the sensitivity that can be obtained in the given observing time. For the correlator output we have retained all 4 polarizations and then assigned polarization data to projects/products as required. We have studied 14 science cases, with the parameters guided by the SKA design reference mission (DRM) document (Lazio, 2011) in most cases.

**2.1 SKA-LOW KSP**

*Probing the Neutral IGM*: This is the EOR experiment in Chapter 2 of the DRM. Frequencies between 50 and 200 MHz (Fiducial Frequency 93MHz) are observed, with channel widths of 150 km/s (10kHz). The required resolution in the image cubes is 120" and only stokes I is retained. In this line case the number of correlator channels is actually set by the bandwidth smearing. 1kHr of observations spread over $2\pi$ steradians is expected to deliver a sensitivity of 10 mJy/channel.

*Low Frequency Continuum:* A generic all-sky continuum imaging experiment at low frequencies. Frequencies between 50 and 300 MHz (Fiducial Frequency 109MHz) are observed. The required resolution in the image cubes is 15" and all four stokes products are retained. 1kHr of observations spread over $2\pi$ steradians is expected to deliver a sensitivity of 1.8 µJy.

*Probing EOR with the Lyman Forest:* This is the EOR experiment in Chapter 4 of the DRM, which covers redshifts of 6 to 20. It is a spectral line, postage stamp, experiment to detect the H$\alpha$ forest at low frequencies. Only one field per beam is expected, on average. Frequencies between 50 and 200 MHz (Fiducial Frequency 93MHz) are observed, with channel widths of 0.2 km/s (which limits the instantaneous bandwidth to be 9MHz). The required resolution in the image cubes is 10" and only stokes I is retained. 1kHr of observations spread over $2\pi$ steradians is expected to deliver a sensitivity of 0.6 mJy/channel.

*HI Absorption:* This is the DRM Chapter 3 KSP to track Galaxy Evolution over redshifts from 0 (1420MHz) to 6 (200MHz) by detecting the absorption line of HI against high red-shift galaxies. It will be a spectral line, postage stamp, experiment. Ten fields per beam are expected, on average. For SKA-LOW this will be at frequencies between 200 and 350 MHz (Fiducial Frequency 261 MHz), with channel widths of 1 km/s. The required resolution in the image cubes is 5" and all four stokes products are retained. 1kHr of observations spread over $2\pi$ steradians is expected to deliver a sensitivity of 0.3 mJy/channel.

**2.2 SKA-MID KSP**

*Galaxy evolution from Nearby HI*: The HI galaxy evolution and large scale structure experiment for redshifts between 0 and 0.1. Frequencies between 1290 and 1420 MHz (Fiducial Frequency 1353 MHz) are observed, with channel widths of 3.3 km/s. The required resolution in the image cubes is 3" and all four stokes products are retained. 10kHr of observations spread over $2\pi$ steradians is expected to deliver a sensitivity of 0.3 mJy/channel.

*Band 3 Continuum:* A generic continuum experiment on SKA-MID. Frequencies between 1650 and 3050 MHz (Fiducial Frequency 2208 MHz) are observed. The required resolution in the image cubes is 0.2" and all four stokes products are retained. 10kHr of observations spread over $2\pi$ steradians is expected to deliver a sensitivity of 3 µJy.

*Pulsar Timing/VLBI:* The KSP pulsar projects are discussed in DRM Chapter 5. Pulsar timing and VLBI are combined as both of these experiments require a number of pointed phased array beams from the correlator (that is the tied array mode) of the inner core. We have allowed for 10 phased beams. Frequencies between 1000 and 3050 MHz (Fiducial Frequency 2208 MHz) are observed, with the output being the time domain signal with two polarisations for VLBI or a filterbank signal with all four stokes products retained. The correlator requirements follow the usual relationships to the inputs but the data product is small. 10kHr of observations spread over $2\pi$ steradians would deliver a sensitivity of 0.9 µJy, but as this experiment would be targeted this estimate is not applicable.

*Pulsar Search:* For pulsar searching the targets are unknown and therefore pointed approaches are not suitable. The requirement is to produced phased array filterbank averages (10MHz and 100 nsec integrations) of the inner core, which cover a larger area of the sky. We have allowed for 100 phased beams. Note that the gains for integrating down to 100 nsec samples and 10MHz channels for a 1GHz bandwidth are minor compared to retaining the Nyquist sampled data. Frequencies between 1000 and 3050 MHz (Fiducial Frequency 1662 MHz) are observed. The correlator requirements follow the usual relationships to the inputs but the data product is small. 10kHr of observations spread evenly over $2\pi$ steradians is expected to deliver a sensitivity of 0.6 µJy.

*HI Absorption:* The KSP to track Galaxy Evolution over redshifts from 0 (1420MHz) to 6 (200MHz) by detecting the absorption line of HI against high red-shift galaxies. It will be a spectral line, postage stamp, experiment. Ten fields per beam are expected, on average. For SKA-MID this will be at frequencies between 350 and 1420 MHz (Fiducial Frequency 578 MHz), with channel widths of 1 km/s. The required resolution in the image cubes is 3" and all four stokes products are retained. 10kHr of observations spread over $2\pi$ steradians is expected to deliver a sensitivity of 0.2 mJy/channel.

*Deep HI:* We have included a deep HI experiment that would aim to directly detect all HI galaxies to much higher redshifts than those in the nearby survey. We have selected a redshift limit of ~1. Frequencies between 650 and 1420 MHz (Fiducial Frequency 937 MHz) are observed, with channel widths of 1 km/s. The required resolution in the image cubes is 3" and only stokes I is retained. 10kHr of observations spread over 200 sq. deg. is expected to deliver a sensitivity of 30 µJy/channel.

## 2.3 SKA-SUR KSP

*Wide and Shallow HI:* We have included a wide (all sky) and shallow (z<0.05) HI experiment such as would be ideal for the SKA-SUR. Frequencies between 1350 and 1420 MHz (Fiducial Frequency 1,384 MHz) are observed, with channel widths of 1 km/s. The required resolution in the image cubes is 10" and only stokes I is retained. 10kHr of observations spread over $2\pi$ steradians is expected to deliver a sensitivity of 0.3 mJy/channel.

*Continuum:* A generic all-sky continuum survey on SKA-SUR. Frequencies between 650 and 1670 MHz (Fiducial Frequency 1005MHz) are observed. The required resolution in the image cubes is 2" and all four stokes products are retained. 10kHr of observations spread over $2\pi$ steradians is expected to deliver a sensitivity of 2 µJy.

*HI Absorption:* The KSP to track Galaxy Evolution over redshifts from 0 (1420MHz) to 6 (200MHz) by detecting the absorption line of HI against high red-shift galaxies. It will be a spectral line, postage stamp, experiment. Ten fields per beam are expected, on average. For SKA-SUR this will be at frequencies between 650 and 1420 MHz (Fiducial Frequency 937 MHz), with channel widths of 1 km/s. The required resolution in the image cubes is 10" and all four stokes products are retained. 10kHr of observations spread over $2\pi$ steradians is expected to deliver a sensitivity of 0.6 mJy/channel.

*Deep HI:* We have included a SKA-SUR deep HI experiment that would aim to directly detect all HI galaxies to much higher redshifts than those in the nearby survey. Frequencies between 650 and 1420 MHz (Fiducial Frequency 937 MHz) are observed, with channel widths of 1 km/s. The required resolution in the image cubes is 10" and only stokes I is retained. 10kHr of observations spread over 200 sq. deg. is expected to deliver a sensitivity of 60 µJy/channel.

## 2.4 The Data Products, Scaling and Processing Power Requirements

The correlator output data rate (both total and baseline averaged) and the accumulated total volume of science data has been computed for each of the 14 survey projects listed in section 2.3. The following table gives example results in Log10 of data product values in Bytes/s or Bytes.

| **Project**       | Cont       | Cont Band 3 | Cont       | Neut IGM      | Deep HI   |
|-------------------|------------|-------------|------------|---------------|-----------|
| **Telescope**     | Low        | Mid         | Survey     | Low           | Survey    |
| Corr. Out (B/s)   | 12.86      | 12.22       | 12.66      | 12.05         | 12.53     |
| Av. Corr. Out (B/s) | 10.52    | 10.24       | 11.3       | 10.88         | 11.85     |
| Sci Image Tot (B) | 15.17      | 15.99       | 18.66      | 13.37         | 18.45     |
| **Project**       | HI Abs     | HI Abs      | HI Abs     | EOR Forest    | HI Deep   |
| **Telescope**     | Low        | Mid         | Survey     | Low           | Mid       |
| Corr. Out (B/s)   | 12.64      | 12.17       | 12.23      | 12.71         | 11.85     |
| Av. Corr. Out (B/s) | 11.48    | 11.03       | 11.54      | 11.54         | 10.71     |
| Sci Image Tot (B) | 12.73      | 13.73       | 15.28      | 11.33         | 16.78     |
| **Project**       | HI Nearby  | HI Wide     | Pulsar Time | Pulsar Search |           |
| **Telescope**     | Mid        | Survey      | Mid        | Mid           |           |
| Corr. Out (B/s)   | 10.78      | 11.36       | 12.79      | 13.79         |           |
| Av. Corr. Out (B/s) | 9.64     | 10.67       | 10.51      | 11.51         |           |
| Sci Image Tot (B) | 13.73      | 16.04       | 13.07      | 18.93         |           |

*Table 2: KSP data rates (Bytes/sec) and volumes (Bytes) for the projects listed in 2.3*

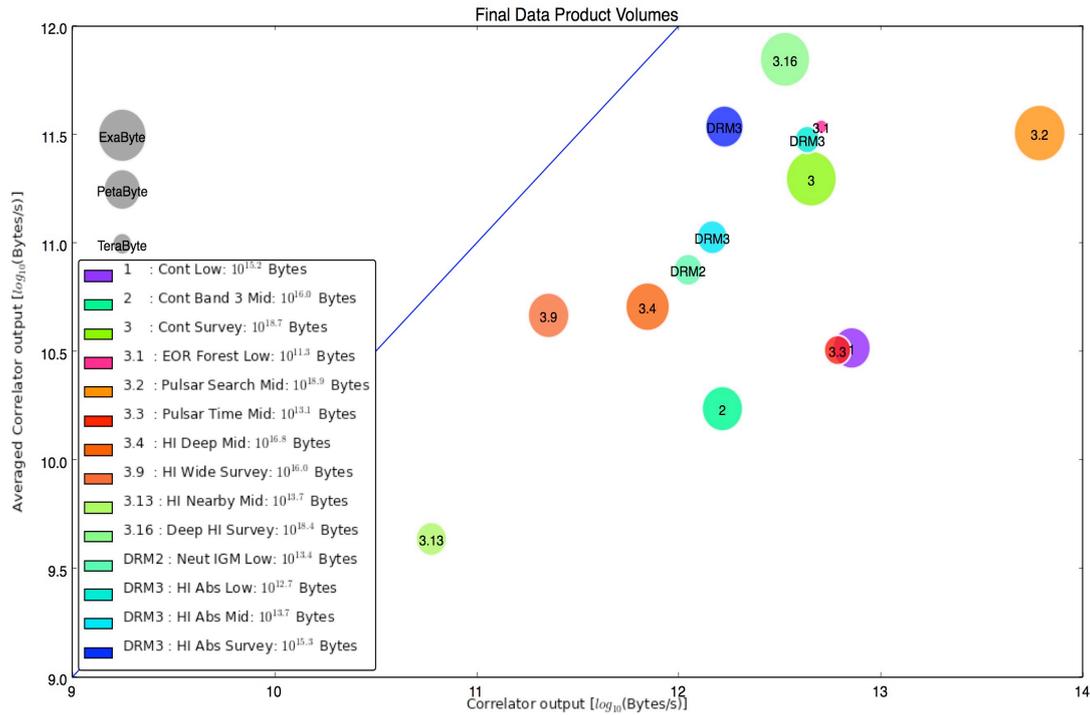

*Figure 1: This diagram shows the optimally baseline Averaged Correlator Output (Level 3 Data) vs. the full, not averaged Correlator Output averaged to the longest baseline (both in Bytes/s) and the Final Data Product (Level 6 Data) volume (in Bytes) for 14 SKA1 KSPs. Note that all three parameters are plotted in Log10 units. The diameter of the circles is scaled by log10 of the total number of bytes of the science data products requested for those KSPs. The scale of the circles is indicated by the grey circles above the legend for Terabytes, Petabytes and Exabytes, respectively. The values are taken from the Table 2.*

For the assumed exposure times and correlator buffer hold time ($10^5$ sec), the distribution of total exposure and data storage requirements for each array is given in Figure 2.

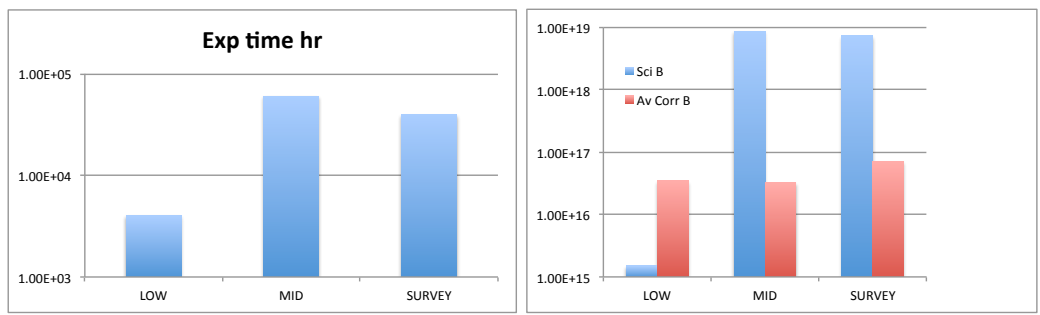

*Figure 2: Total exposure time in Hours (left panel) and total science data volumes (Bytes) for the KPSs on the LOW, MID and SURVEY arrays. Correlator sizes referred to averaged outputs for the largest project on each array and $10^5$ second hold times.*

These results scale following simple relationships under the assumption of fixed maximum baseline length. Increasing the number of antennas increases the raw input by the same factor and the correlator output by that factor squared. It has no effect on the science data product, given a fixed observation time, as the number of visibilities that form the image improves the sensitivity but not the size of that image. Increasing the observing time does not change the correlator inputs or outputs, but directly increases the size of the final data product by the same factor. Not included in this analysis, because of our starting assumptions, is that increasing the number of antennas and thereby increasing the instantaneous collecting area would reduce the required observing time by the same factor. This in turn would reduce the data product size. That is, if costs were dominated by storage factors, increasing the number of antennas reduces the overall storage costs, provided the science goals are not changed. These relationship are shown in Table 3. It is important to note that the cost of servicing a particular data rate or data volume can be switched between capital construction costs (size of storage facility) and operational costs (number of days of operations needed to deliver the KSP) depending on what observational parameters are held fixed.

| Data | Exposure time fixed (**T**) | Number of Antenna fixed (**N**) | NT fixed: **NT=x** |
|---|---|---|---|
| Raw (B/s) | $N^1$ | $T^0$ | $xT^{-1}$ |
| Corr (B/s) | $N^2$ | $T^0$ | $x^2T^{-2}$ |
| Products | $N^0$ | $T^1$ | $xN^{-1}$ |

*Table 3: The dependence of the Raw and Correlator data rates and the final data product size, on the number of antenna (N) and the exposure time (T) assuming fixed N, fixed T or fixed sensitivity measure (x) under **the assumption of a fixed maximum baseline**. Green indicates a construction cost driver and red indicates an operations cost driver.*

Current radio astronomy calibration algorithms are iterative and thus require keeping all the data of an observation on a buffer storage and allowing for several reads over the whole data set. Given the estimated input data rates in the table above, this translates into several 100 PB storage buffers with a sustained random I/O speed of approximately 30 TB/s, for an observation duration of 12h.

The computational complexity (i.e. operations/Byte) in radio astronomy processing is typically fairly low and that does not fit well with current computing architectures. The expected number of real (delivered to applications) FLOPs required is of order 100 PetaFLOPs across all three telescopes. In addition, the efficiency of typical current day super-computers is optimised towards very high Linpack performance, but Radio Astronomy is dominated by FFTs and the efficiency is far lower. The average efficiency shown in Figure 3 below is just 6.9% (Wu, 2014). ***This means that we would need either super-computers, which are optimised for our compute tasks, or we would need super-computers with a Linpack performance (in FLOPS) about 10 times higher than our actual compute requirements***. The latter would imply much higher capital and operational (power) costs.

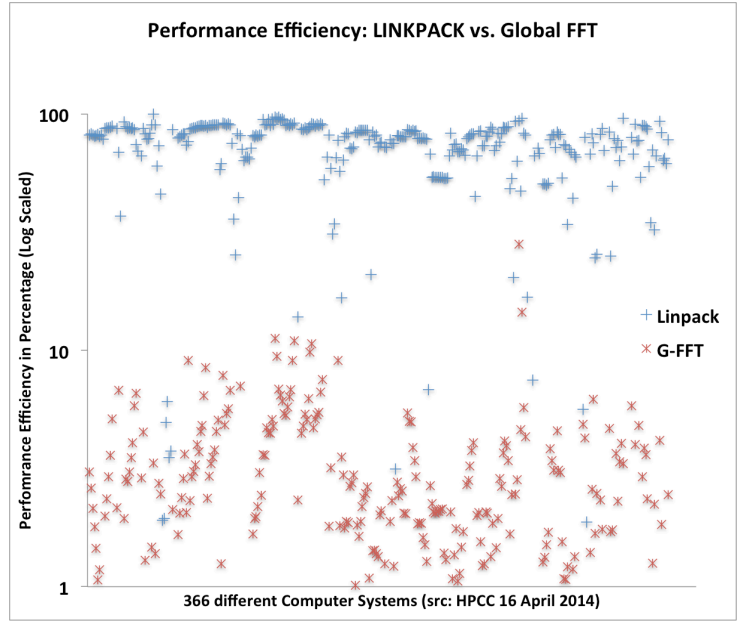

*Figure 3: Performance efficiency of 366 different HPC systems. Blue crosses are Linpack, red stars are FFT. The average FFT efficiency is just 6.9%. [1]*

## 3. SKA Data Flow – construction and operations

How will the SKA manage the flow of these data volumes through the Observatory? Where will the flow of data start and stop and what will be the resultant construction and operational consequences for the SKA project? Who has responsibility for the various processes and products involved in doing SKA science? In this section we will outline a likely model of the SKA Observatory data and process flow in order to identify the interfaces and processes that need to be scaled and costed for ExB astronomy. We will then analyse the scale of operational and construction requirements based on the data requirements of the SKA KSPs.

### 3.1 The SKA Observatory and Data Flow

The international SKA Organization (SKAO) is now working towards the creation of telescopes and observing facilities in Australia and South Africa as part of SKA Phase 1. With the SKAO headquarters at JBO, these facilities will form the SKA Observatory that will deliver a ground-breaking observational capabilities for global astronomy in the second decade of this century. Like many transformational observatories of the past, the SKA Observatory will consist of complementary capabilities in frequency and sky coverage that enable a range of survey and pointed programs to the depth afforded by the square kilometre aperture.

While the planning for the technical and scientific operations of the SKA is still to be finalized, the SKA Observatory will most probably execute processes and generate products that are similar to other successful international "public" observatories on the ground and in space. Over the past 20+ years, a similar science operational paradigm has been adopted by optical/IR (e.g. Hubble, VLT, Gemini, CFHT) and radio (e.g. ALMA) observatories as documented in the ongoing SPIE series on Observatory Operations (Ed. Quinn; 1998, 2000, 2002). Within this end-to-end paradigm, the Observatory takes responsibility to accept and adjudicate proposals, schedule resource, collect data, provide calibrations, exercise quality control and deliver data packages (see

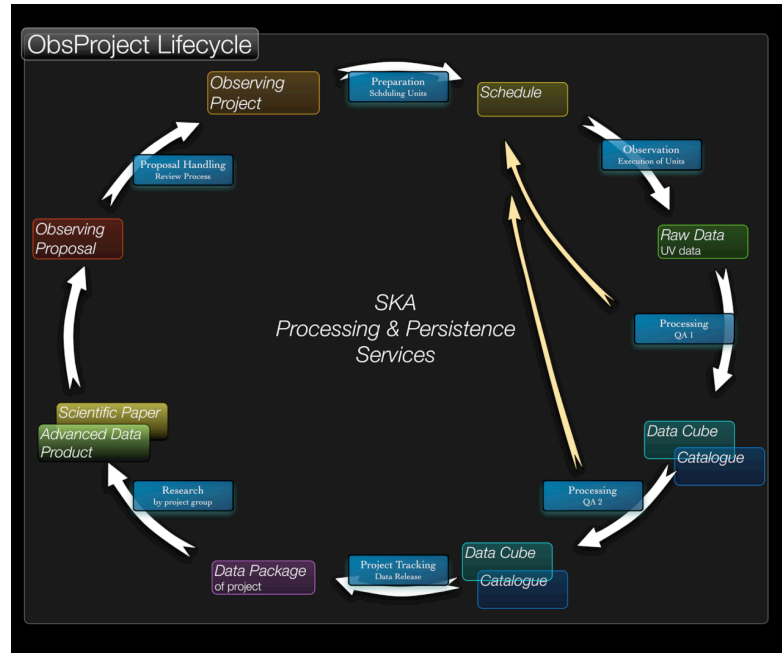

Figure 4).

*Figure 4: Possible SKA end-to-end process and data flow (the "egg") based on similar systems implemented at ESO and elsewhere.*

These delivered data packages could be rudimentary or advanced depending on the obligations and resources of the Observatory as well as the agreed boundary line between the Observatory and the community in terms of the total effort required to deliver scientific results. Where is this boundary for the SKA Observatory? What does the Observatory deliver and what does the community have to undertake? How should it be defined? Is it different for different types of programs (e.g. survey verses pointed verses target of opportunity)? These are clearly fundamental questions that need to be addressed as we plan the construction and operations of the SKA.

The flow of processes and products surrounding the lifecycle of an SKA project can be divided into 4 domains.

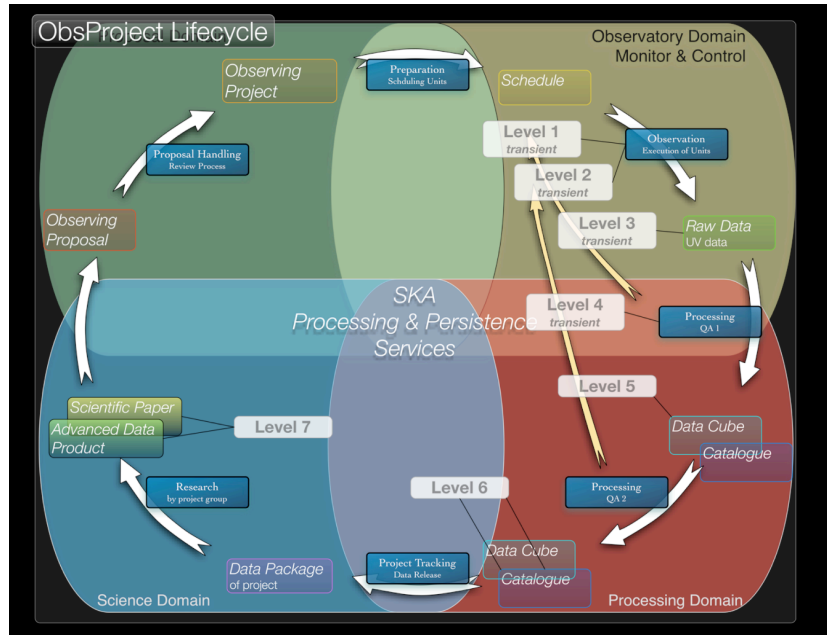

*Figure 5: Operational domains and associated levels of data produced – see Figure 6*

In the Proposal Domain proposals are prepared by the community, submitted and judged by the Observatory TAC process leading to the award of time and the preparation of a project description (sometimes called Phase 2 Proposal Preparation) in a form that can be scheduled and executed. In the Observatory Domain the observations are scheduled, executed and raw data is generated and captured. The Processing Domain turns raw data into images and catalogs and involves calibration and quality control at multiple stages. Finally the Science Domain defines and delivers data to users and facilitates the creation of advanced data products from input cubes and catalogs. The Observatory and Processing Domains are clearly central to the functioning of the SKA and will be assumed from here on to be part of the construction and operations cost of the SKA. Proposals are clearly defined and submitted by the community and the community will be doing science with SKA released data cubes and catalogs. Under these assumptions, the SKA will need to manage and deliver a number of different data products specified as data levels. These levels are shown in Figure 6.

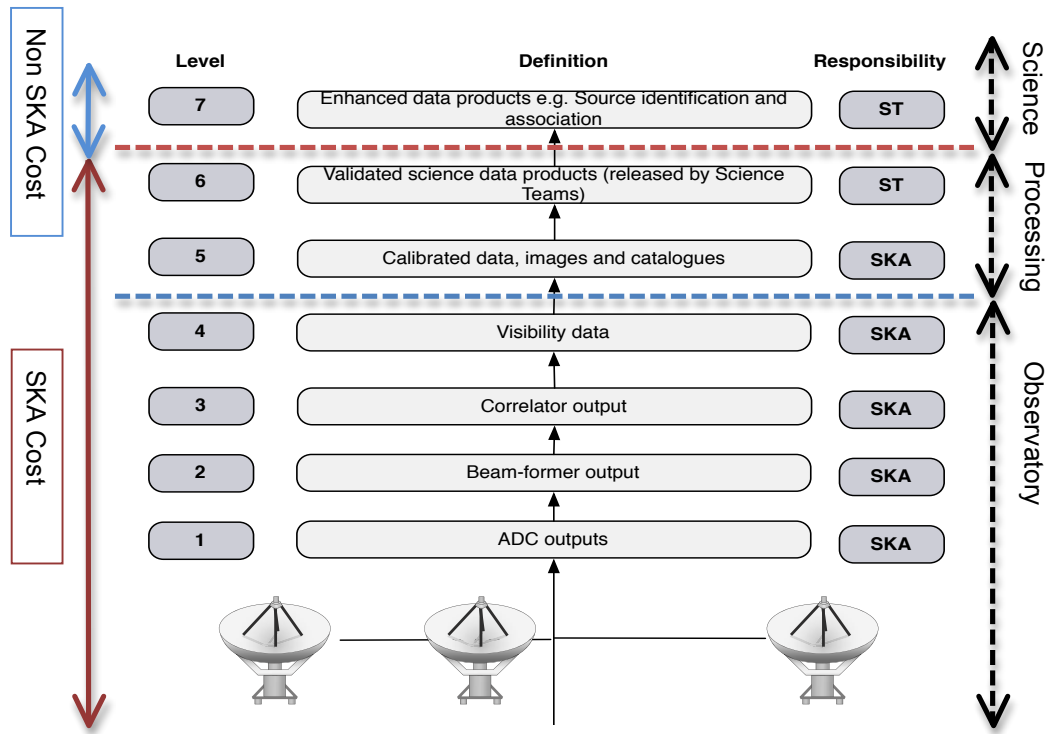

*Figure 6: SKA data product levels (shown in Figure 5 from Cornwall, 2013) The blue and red dashed lines show the boundaries of the Observatory, Processing and Science Domains of Figure 5. The SKA Organization currently assigns responsibility to the KSP science teams (ST) to validate products to Level 6 and initiate the development of enhanced data products (Level 7).*

At this point in time, the SKA Organization has indicated a transition from SKA Observatory responsibility for the data flow, to community responsibility between Level 5 and Level 6 data products. Whether this change of responsibility involves the utilization of SKA Observatory resources by the community to execute a quality control process and create validated products at Level 6, or whether it requires additional resources, is not determined at this time. The additional resources needed for the creation of Level 6 products will most likely be small compared to those required for Level 5. For this study, we will assume the Level 5 and 6 resources are within the construction and operations budgets of the SKA Observatory.

### 3.2 Construction and Operations Resources for Level 1 to Level 6

Based on the data volumes and rates in Table 2 for the assumed exposure time and telescope, Figure 7 shows the predicted volume of data products (Level 6) and Corrleator output buffer sizes assuming a buffer hold time of 100,000 seconds.

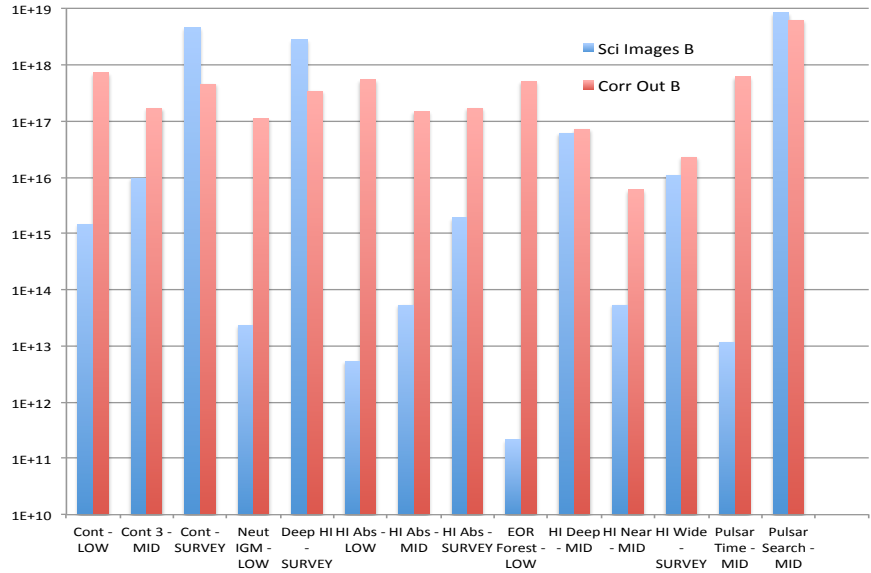

*Figure 7: Total bytes generated at Level 6 for each KSP and the associated Correlator buffer size (without averaging) with a buffer hold time of $10^5$ seconds.*

If all KSPs were completed and the Level 6 products held on storage, the total science storage volume would be **16 Exabytes** with the largest buffer being **6.2 Exabytes** required by the Pulsar Search (MID) KSP. Figure 8 shows the effect of using averaged correlator outputs. In this case the required maximum buffer size is **71 PBytes** for the Deep HI Survey (SURVEY).

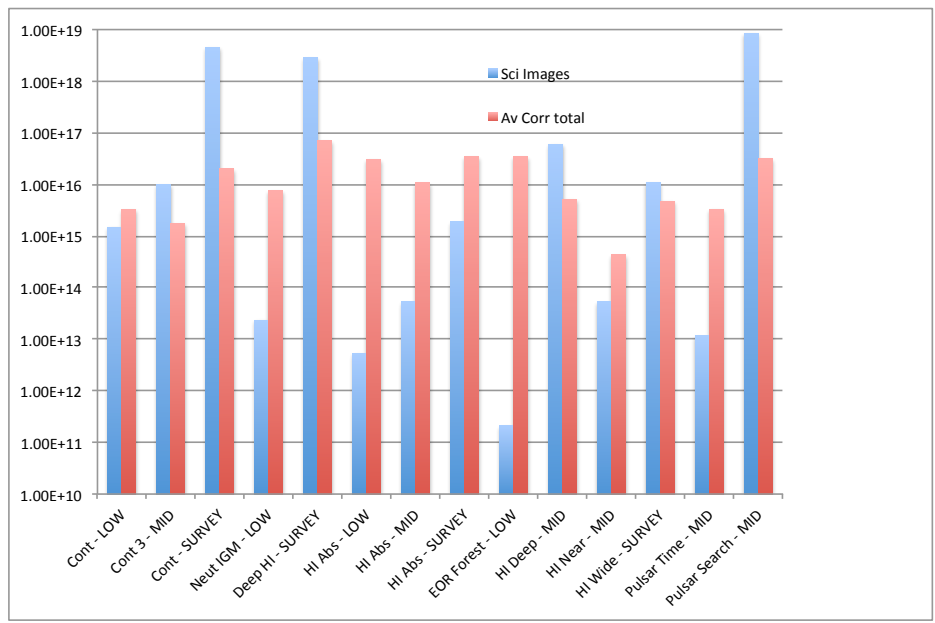

*Figure 8: Total bytes generated at Level 6 for each KSP and the associated Correlator buffer size (averaged) with a buffer hold time of $10^5$ seconds.*

The approximate cost of storage can be estimated from recently constructed HPC centres (e.g. Pawsey Centre in Australia) and from published industry trends for disks, tape robotics and hierarchical storage management systems given in the table below.

| Year | $ HPC Doubling Exponent (yrs) | $ Tape Media Doubling Exponent (yrs) | $ Disk Doubling Exponent (yrs) | Tape media cost $/GB | Total Tape System $/GB | Disk $/GB |
|---|---|---|---|---|---|---|
| 2014 | -1.5 | -2.3 | -3 | 0.06 | 0.3 | 0.07 |

*Table 4: Current trends (doubling exponents) and media costs for HPC, disk and tape systems based on HPC centre construction costs and industry trends.*

Based on this analysis, and assuming correlator averaging, the storage costs on the timescale of 2020 are dominated by the tape storage systems (not correlator buffers) for the three largest surveys (Continuum, Deep HI and Pulsar Search) each of which will require investments of order 100 MEuro. The associated HPC cost (assume 100 PFlop off-the-shelf performance) will be approximately 80 MEuro by 2020.

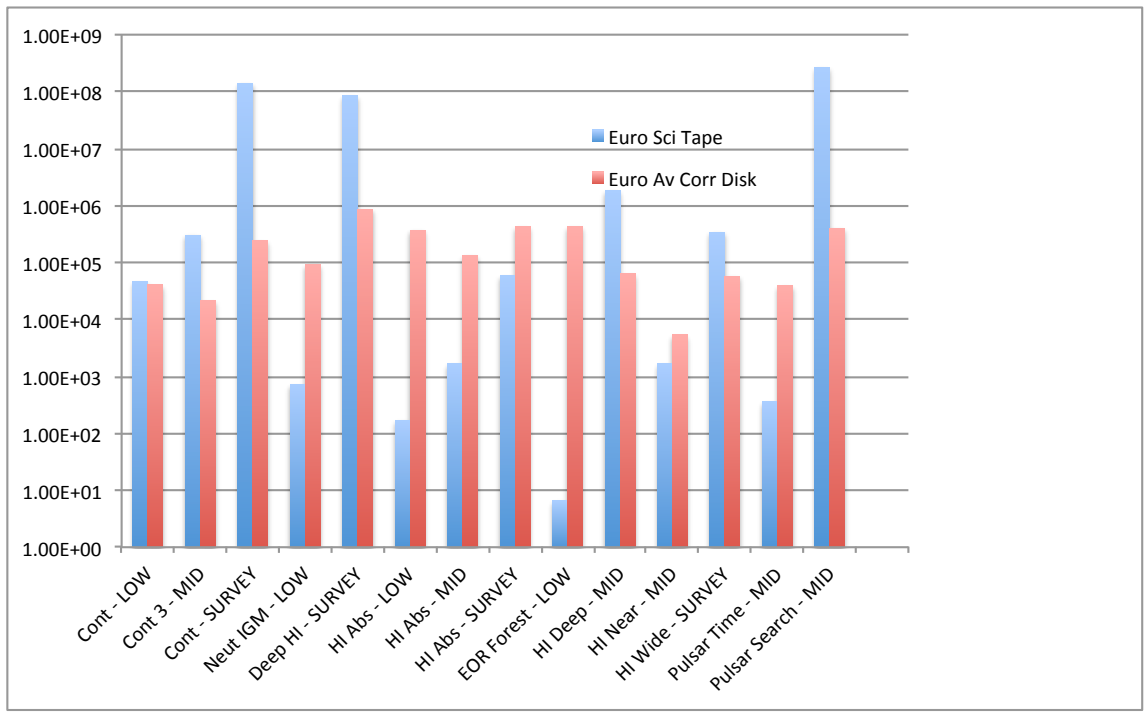

*Figure 9: The cost in Euros of tape and disk storage systems for each of the KSPs projected to 2020 using the parameters in table 4.*

The operational costs associated with these storage systems, along with the associated HPC, can also be estimated based on the recent HPC centre completion costs and industry trends in the power performance of HPC. Taking the power cost as a **lower limit** to the operational cost and assuming a Green500 mean doubling timescale for improvement in watts/mflop of 1.6 years, the lower limit operational cost of a 100

PFlop facility is approximately 30 MEuro/yr on the timescale of 2020 at fixed unit electrical price. Figure 10 shows the construction cost and operational cost (over the total exposure time) for the combined set of KSPs as well as the three largest surveys under the assumptions above. The Figure also indicates the expected operational and construction costs for SKA1 based on the 650M Euro cost cap, a 100 MEuro allocation to HPC/Storage/Software, a total operational budget of 65 MEuro split equally among the three sites (UK, RSA and AUS) with a 50:50 split of operational costs between a site and a HPC/Data Centre.

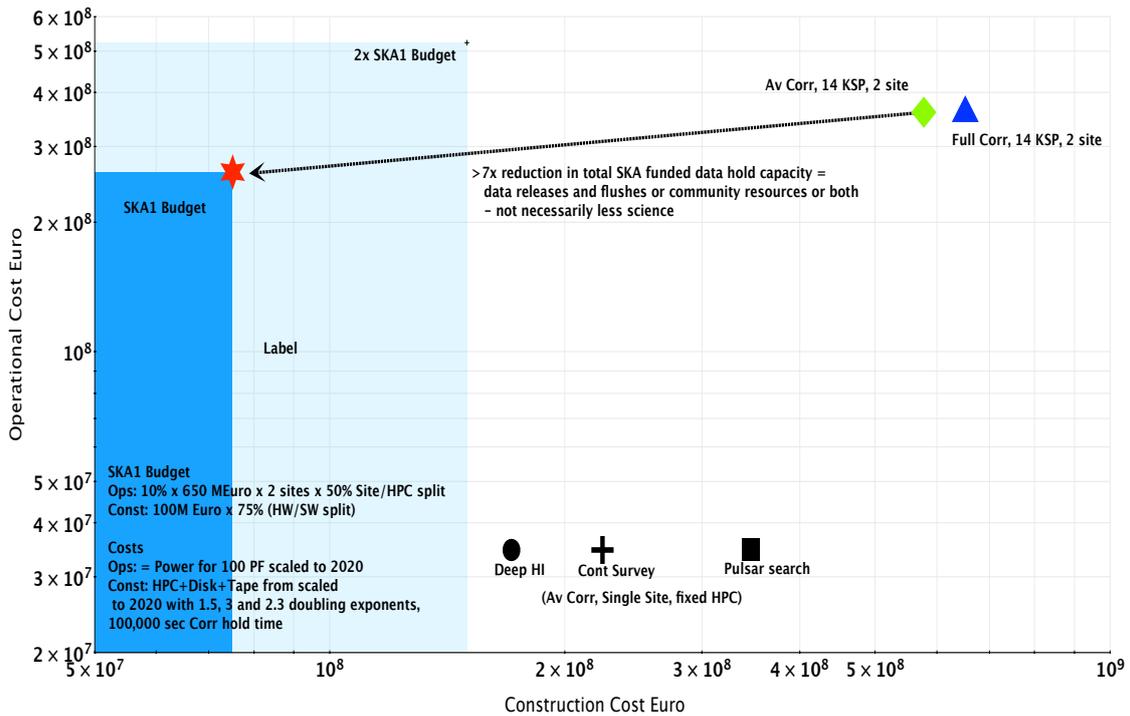

*Figure 10: Projected operational and construction costs for KSP projects over the required exposure time, compared to estimated SKA1 budgets.*

From Figure 10 it is apparent that the scale of the storage and processing/power requirements for SKA1 KSPs will exceed SKA1 expected budgets by a factor of at least 7 given the operational costs are lower limits. If the scale of the science requirements remains unchanged (survey sensitivity/depth) then the SKA funded data holding capacity needs to be reduced by at least the same factor of 7, or more, for Level 6 products. This could be achieved by instituting 7 or more data releases in the life cycle of a survey with a subsequence flush/reuse of the SKA funded resources. This would place the long term storage responsibility for SKA science data into community hands. Furthermore, the resources needed for Level 7 science are clearly of the same order, or larger than, Level 6 requirements. We are therefore facing a situation in which a significant community commitment, outside of the funding to construct and operation SKA1, will be required to deliver SKA1 science. This situation has been faced before by projects like the Large Hadron Collider and there is now clearly a need to initiate a significant effort in planning and resourcing the delivery of SKA science by the SKA community, working closely with SKAO.

## 4. The SDSS, LSST and LHC Science Data Systems

As pointed out in Section 3, the delivery of SKA science is going to require an SKA-Data system that is focused on delivering science from surveys in the most efficient manner possible and which is a distributed and shared facility across the SKA Observatory and the community. In the past 20 years, there are several important projects from astronomy and particle physics that have shared the mission of delivering maximal science impact to a distributed community within limited construction and operations resources. In this section we will examine three such projects. In each case we will look at the approach, the costs/challenges and whether these approaches will scale to the SKA needs.

### 4.1 The Sloan Digital Sky Survey (SDSS)

The SDSS (www.sdss.org) has made a significant contribution to international astronomy over the past 10+ years. This scientific return has resulted primarily from the development of a specific database and data management system that allowed the community to launch queries on a final set of catalogs and images (Level 6). The SDSS system is therefore a role model for the types of facilities we would expect to see in the Science Domain of SKA and dealing with the generation of Level 7 and beyond products.

#### 4.1.1 Approach

The group at the Johns Hopkins University (Alex Szalay and Ani Thakar, in collaboration with Jim Gray) have spent much of the last two decades working on the archive for the SDSS. Opening in 2001, the archive has been issuing new data releases yearly. The most recent version of the database has a 12TB queryable public core, with about 120TB additional raw and calibrated files. The collaborative CasJobs interface has more than 6,820 registered users − almost half of the professional astronomy community.

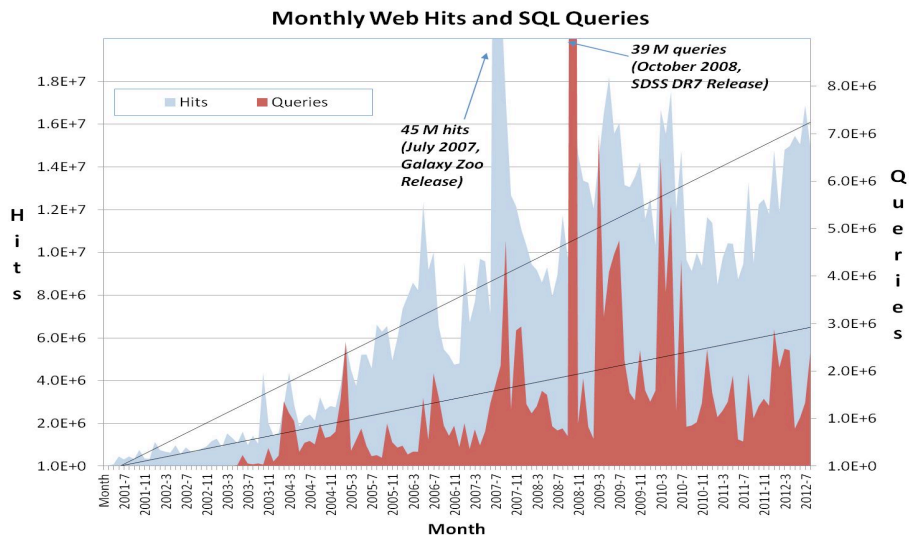

*Figure 11: Monthly traffic of the SDSS SkyServer [provided by SDSS]*

SDSS (2000-2005) and its successors SDSS-II (2005-2008) and SDSS-III (2008-2014) surveys have produced data to support 5,000 papers with more than 200,000 citations. Within the Collaboration there have been over 120 SDSS-based PhD theses, and outside the Collaboration there have been many more.

### 4.1.2 Challenges and Costs

As traffic on the SDSS archive grew, many users were running repeated queries extracting a few million rows of data. The DB server delivered such data sets in 10 sec, but it took several minutes to transmit the data through the slow wide-area networks. It was realized that if users had their own databases at the server, then the query outputs could go through a high-speed connection, directly into their local databases, improving system throughput by a factor of 10. During the same time, typical query execution times and result sets kept growing, thus synchronous, browser-based access was no longer enough, especially for the rapidly growing segment of "power users". There had to be an asynchronous (batch) mode that enabled queries to be queued for execution and results to be retrieved later at will.

The CasJobs/MyDB batch query workbench environment was born as a result of combining these "take the analysis to the data" and asynchronous query execution concepts. CasJobs builds a flexible shell on top of the large SDSS database. Users are able to conduct sophisticated database operations within their own space: they can create new tables, perform joins with the main DB, write their own functions, upload their own tables, and extract existing value-added data sets that they can take to their home environment, to be used with the familiar tools they have been using since their graduate years. Their data and query history is always available to them through their user id, and they do not need to remember things like "how did I create this data set?"

As users became familiar with the system, there were requests for data sharing. As a result, the ability to create groups and to make individual tables accessible to certain groups was added. This led to a natural self-organization, as groups working on a collaborative research project used this environment to explore and build their final, value-added data for eventual publication. GalaxyZoo (www.galaxyzoo.org), which classified over a million SDSS galaxies though a user community of 300,000, used CasJobs to make the final results world-visible, and CasJobs also became a de-facto platform for publishing data.

An additional challenge was in dealing with regular data releases, in essence a versioning problem. As the project moved along, data was constantly collected, processed and ingested into an internal database. However, it would have been very confusing for the public to access this database, as the same query could have easily returned a different result every day, due to changes in the underlying data. Thus the project adapted a yearly data release. These happened usually coincident with the summer meeting of the American Astronomical Society. Data releases were accompanied by a short paper in the Astronomical Journal, which made citations to the data particularly easy through a conventional mechanism.

Older data releases were still kept alive, even today, as a research project that has been started with a given data release would have to rely on consistency and reproducibility. It is best to think of the Data Releases as ever newer editions of a book, where we do not throw away the old ones, they are still kept on the bookshelf.

The original costs of the data infrastructure for SDSS have been grossly underestimated. The original budget for the whole project was about $25M in 1992, of which about **$2M** was the projection for the software and the archive. The estimated raw data was expected to be around 10TB, with a 0.5TB catalog database. Much of the software was going to be recycled from existing modules and data management tools from high energy physics. It was gradually realized that all of these costs and data sizes were highly underestimated. In the end, the project took more than 16 years to complete, and the total cost exceeded $100M, of which the software, the computational hardware and the archive were approximately 30%.

|  | **Raw Data** | **Database** | **Project Cost** | **HW/SW cost** |
|---|---|---|---|---|
| **Initial** | 10 TB | 0.5 TB | $25M | $2M |
| **Current** | 120 TB | 12 TB | $100M | $30M |

*Table 5: Predicted and actual costs and sizes for SDSS data systems*

However, as the final data sets for the original SDSS-I and SDSS-II surveys are archived for the long term, it is clear that the data that needs to be stored is at least 120TB, the main database is over 12TB, with 25TB of additional databases (Stripe 82, Target, etc). So it is fair to say that the final data products were considerably more ambitious than the original vision. Nevertheless, the task was made easier by the exponentials: Moore's Law, and Kryder's Law. Both the speed of computing at a fixed yearly cost, and the capacity of disks at a fixed budget was growing exponentially, and the delays in the project enabled a much more ambitious data management outcome.

### 4.1.3 Scaling to ExBs

The current system with its 12TB database does not need a distributed system, so taken directly it would not scale to even a Petabyte. However, the different modifications of the SkyServer are now approaching a Petabyte in a single database, but spread over many servers, similar to the shared approach of LSST. If there was a lot of time pressure, one could build a 10PB astronomy database based on today's version of the SDSS software without much difficulty, probably at a cost of a few $M.

### 4.2 The Large Synoptic Survey Telescope (LSST)

The LSST is the highest priority of the US Decadal Survey of Astronomy for the period from 2010-2020. It will be an unprecedented optical survey facility capable of imaging the entire hemisphere every three nights and producing catalogs with billions of entries. The program of the LSST utilizes one telescope and one instrument to execute a number of well-defined surveys with community science focused around released catalogs and images. It is therefore has a more specialized approach to dataflow than the ESO-like system and is conceptually similar to SDSS.

### 4.2.1 Approach

LSST is a wide-field imaging survey that repeatedly images about 20000 sq deg of sky in 6 broadband filters over a 10 year period. Individual images are 3.5 gigapixels, and are collected at roughly 20 sec intervals. While LSST will be used to address a wide variety of science areas, there is a single survey with a uniform set of data products which will be used for all of them. In many ways LSST can be thought of as a mega-SDSS, although without a spectroscopy component. Its major differences from SDSS are a much greater image depth (single exposures reach approximately 24.5 mag in r-band), and an emphasis on time dependent phenomena.

As was the case with SDSS, LSST expects that most science will be done with catalogs produced by the data pipelines, rather than directly from the images. The catalogs are implemented as a relational database, and one of the major design challenges of the data management system has been to ensure that this database is scalable to a large number of nodes that efficiently process database queries in parallel. This is required to support database queries that scan the entire database, as is required for many science use cases. A second challenge is to design the data processing pipelines so that they produce the catalog quantities that are required by the science use cases, and that they do so with a verified data quality that makes it unnecessary for science users to recompute them on their own from the image data.

LSST's image storage approach leverages the fact that most science that does require going beyond the catalogs to the actual images will do so by processing the deep coadds rather than individual exposures of a field. An example is searching for gravitationally lensed galaxies. Current processing pipelines are unable to reliably identify gravitational lenses, so images must be processed by a combination of specialized catalog-based selection in conjunction with human eyes. These lenses are static on LSST time scales, however, and the science need is for maximum depth and resolution, so the coadds will be utilized.

Since LSST images a typical field 200 times in each band, the coadds account for only about 0.5 percent of the total image data. This allows the vast majority of the image data to be archived on tape, which is inexpensive but slow compared to rotating disk. The typical image, therefore, is accessed only by the pipelines during the production of a data release, of which there are a total of 11 during the survey. The rest of the time, it can quietly reside on tape, available if needed, but with the expectation that it will not be.

**4.2.2 Challenges and Costs**

In considering the scalability of LSST's data management system to SKA size, we begin with a simple estimate of how big the scaling factor is. The overall size of the LSST data products is within an order of magnitude of SKA's. At the end of its 10 year survey, LSST will have archived roughly 500 PB of images. This is comparable to the deep HI survey, and roughly one tenth the size of the continuum survey, which together dominate the overall size of the SKA data products. The size of the LSST catalogs will certainly be substantially larger than SKA's, simply because of the difference in the total number of objects detected (roughly 45 billion objects for LSST vs 10s of millions for SKA). Considering only storage, the LSST's infrastructure cost is estimated at $6.3M and its operating costs (mainly in purchase of tapes and replacement disks) is $1.3M/yr. These costs are in FY15 year dollars, with projected component costs appropriate for operations beginning in 2022. This would suggest an SKA cost of perhaps $60M and $10M/year for data systems alone. These numbers are consistent with values in Figures 8 and 9.

#### 4.2.3 Scaling to ExBs

From that simple perspective, then, it would seem that LSST's approach could scale to the SKA without unreasonable cost or difficulty. In fact, however, this scaling argument is of limited value because the SKA's will use its data in a very different pattern than does LSST. The patterns differ in at least two ways. First, while most LSST science will be based on catalog data, SKA science will mostly directly access images. Second, while most LSST image access outside of the pipelines themselves will be to a very limited subset (the deep coadds), the SKA will require much more uniform access to images. This is in good part a reflection of the fact that the typical SKA "image" is a three dimensional image cube of substantial depth, while an LSST "image" is basically two dimensional (or three dimensional with negligible depth to cover the six filter bands).

These considerations suggest that the LSST's approach of keeping the majority of its images stored on tape will not be practical for the SKA, and that the consequent cost of the required SKA infrastructure will be much larger than implied by the simple scaling argument.

### 4.3 The Large Hadron Collider (LHC)

The LHC is an international facility that was one of the first science endeavours to venture into the PetaScale of data and processing power. In facing this challenge, the LHC has split its data flow, responsibilities and costs into a number of tiers. A similar structure may be essential and desirable for SKA, for a number of community and cost-cap related reasons.

#### 4.3.1 Approach

There are four large detectors at the LHC, which in the first 3 years of LHC running (Run 1) produced around 70 PB of raw data between them. This is significantly more than the 15 PB/year originally anticipated. Over the next 10 years the LHC and the experiments will all see significant upgrades, increasing luminosity and resolution, resulting in significantly increased data rates and volumes. For the "High Luminosity LHC" (HL-LHC) starting in 2023 or so, some 400 PB/year of raw data are anticipated; already in LHC Run 2 (2015-2018) double the Run 1 data volumes are expected, with intermediate increases for the run in 2019-22. Although today's raw data is of order 25 PB/year, the total data quantities are much larger; taking into account the needs of simulation, the various processing passes, and analysis data, the large experiments (ATLAS and CMS) each have a managed dataset of the order of 200 PB. Thus on the HL-LHC timescale – very close to the SKA timescale – each of ATLAS and CMS are likely to have (multi-)Exabyte scale datasets.

When the planning for LHC computing began in earnest around 2001, it was clear that the cost of computing, storage, and networking for LHC would be significant, and would approach the cost of building the large detectors. Thus CERN decided that the LHC computing project would be treated in terms of scientific and funding review, in the same way that the large experiments are managed. ***Thus, for the first time in High Energy Physics, the computing was recognized as one of the key components of enabling the science, together with the accelerator and the detectors***. *It was also clear* *that the only way that sufficient computing and storage resources would be available would be by direct contributions from the countries participating in LHC experiments.* It was also clear that those contributions would be through computing infrastructures in the funding countries. This funding reality then drove the need for building a globally distributed computing infrastructure.

The Worldwide Computing Grid (WLCG) collaboration was created to manage this computing infrastructure, and the reporting and reviewing requirements. It is a collaboration bound through a Memorandum of Understanding (MoU) between CERN and each of the participating funding agencies. This is the model of the LHC experiment collaborations themselves. An important feature of this MoU is that this is the mechanism through which the funding agencies pledge funding for the computing and storage resources. This is done through a yearly review and pledging process. It should be noted that the pledges are in terms of resources and not as money. As part of the MoU the computer centres agree to provide the services and service levels defined for the various Tiers.

High Energy Physics (HEP), and thus, LHC data processing and analysis differs from astronomy in several ways. In HEP there is no concept of a "science data product" in the sense of the output of an observatory (or a LHC detector). The output of the detectors is raw data, which must go through a series of data reconstruction, processing and selection stages, producing condensed summary data that is used for physics analysis. However, there may be different such analysis data sets, depending on the types of analysis to be performed. All of these steps are performed on the globally distributed computing infrastructure (the WLCG). In addition, a significant amount of simulation is needed, recognized as part of the analysis chain, and also performed on the distributed infrastructure. CERN, as the accelerator laboratory has a special role in that it archives on tape all of the raw data, and performs the necessary quasi-online calibration tasks. In terms of overall capacity CERN provides between 15-20% of computing and active storage needs of the experiments.

The computing model adopted for all LHC experiments initially is the following. All raw data is archived at CERN (the Tier 0), and a second copy distributed between 11 large (Tier 1) computing centres worldwide (7 in Europe, 2 in USA, 1 in Canada, and 1 in Taipei). The Tier 0 provides first-pass data reconstruction (particle track-finding), and urgent calibration tasks.

The Tier 1 centres provide reprocessing of the raw data once the calibrations have improved, creating the analysis data sets. These are distributed to the Tier 2 analysis centres.

The many Tier 2 centres (today ~150), often University Physics departments, provide compute clusters for physics analysis, with appropriate amounts of storage to host the physics data sets. All collaborators in an experiment have the right to access resources at any site, and indeed should not need to know where the data and compute is located.

There are also local resources available to national groups, out of scope of the WLCG collaboration.

### 4.3.2 Challenges and Costs

One of the early concerns was the adequacy of the networking, in particular being able to ensure the rapid distribution of raw data to the Tier 1s to ensure data safety. This was addressed by provisioning dedicated 10 Gb/s connections from CERN to each Tier 1. This rapidly evolved to a redundant set of connections with backup paths between Tier 1s. This is the LHC Optical Private Network (LHCOPN), funded primarily by the Tier 1s and provisioned by the NRENs and Geant (in Europe), and a DOE-funded project in the USA. This was often seen by the NRENs as an important thing to help with as the LHC science was so visible. The costs of building LHCOPN were well within the computing budgets of the Tier 1s.

      The experience in the run-up to LHC start, and the subsequent 3 years of Run 1 is that the network not only performs as needed, but is in fact a resource that can be used to reduce the eventual storage costs. This has led to an evolution of the strict hierarchical model, with data moving between sites as peers, and networking developing to allow this. The LHC experience, and the rapid evolution of network technology lead us to believe that over the next 10 years this will not be a technology concern, and that on the timescale of HL-LHC and SKA 1-10 Tbps networks will be available and affordable.

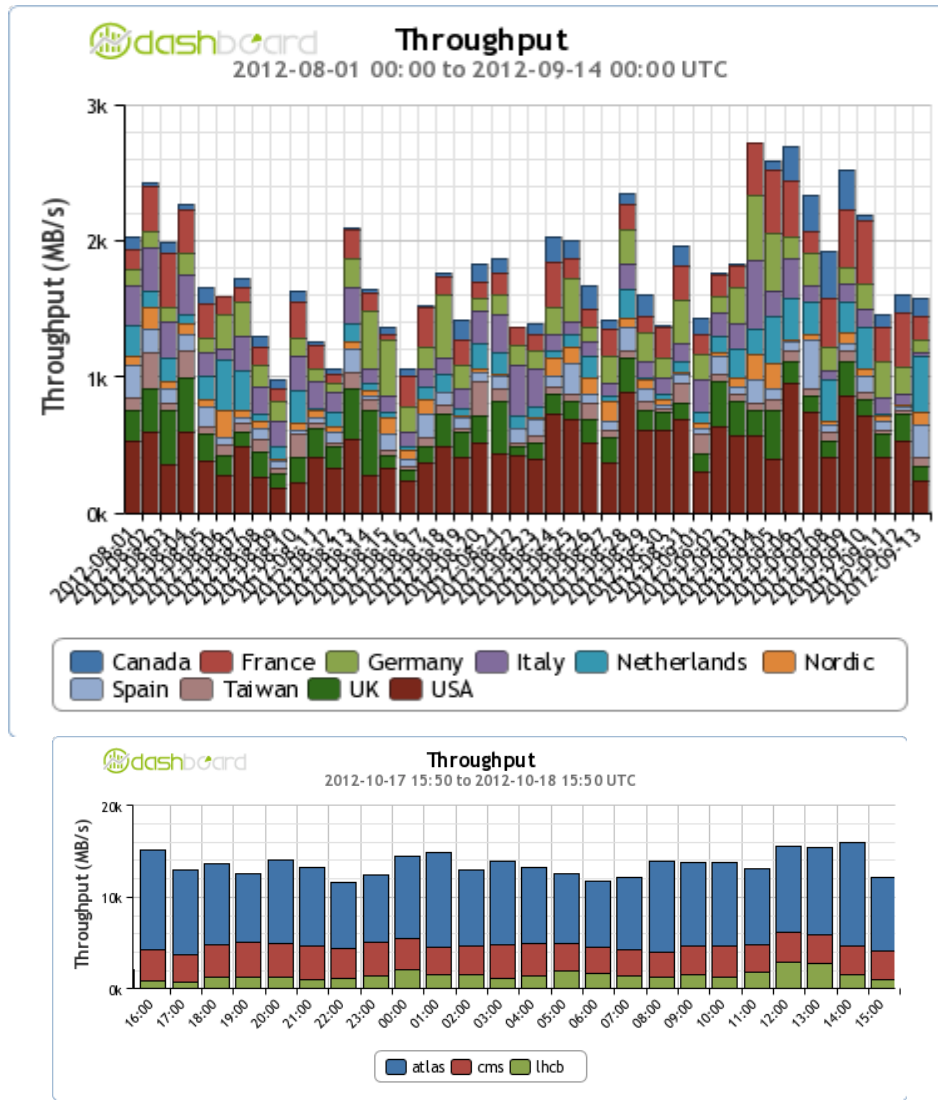

*Figure 12: Examples of sustained network performance for LHC: top: 2 GB/s data export from CERN; bottom: global data transfers at 15 GB/s [data provided by CERN]*

Today LHC spends around 60% of its computing budget globally on disk storage, a lot of this driven by the distributed nature, and the need to have many data copies to be able to utilize available compute resources. In the last year or so, the experiments have started to make use of accessing data remotely: pre-placing it "just-in-time", fetching it from a remote site if needed and not available locally, and in some cases realtime remote I/O. These strategies can be used in different cases, based on the observation that moving data today is cheaper than storing it. Thus, rather than storing many copies, there will many more network accesses to fewer copies of the data. This is at an early stage, but is likely to be the basis of the computing models in the coming years.

**4.3.3 Scaling to ExBs**

In the longer term – on the SKA timescale, this computing model must evolve significantly to make better use of economies of scale. It is clear that a highly distributed system such as the WLCG is more expensive than a more centralized large-scale system could be. Until now, this consideration has been outweighed by the considerable advantage of funding national institutions to provide the resources, allowing a real sense of ownership of the data, and complete engagement of the physicists in the analysis. Particularly for institutes outside of Europe and North America, this aspect has been essential, as has been the aspect of hands-on training of young scientists in managing large scale data and the associated technologies. In the future, however, we must make much more effective use of the available funding, building on technology evolution to federate national computing structures as they also evolve. It is also clear that on this timescale the actual computing models must evolve, to reduce the overall amount of processing necessary, making early decisions about which data to keep and which to reject. There are also ideas following the astronomy model of better defining the boundary between processing to produce a science data product and the use of those to do science. It is clear that the system will be highly distributed but that the model must evolve to better hide that nature from the scientist.

**5. Delivering SKA science 2020+**

The analysis of the scope and cost of SKA1 KSPs in sections 2 and 3, and the experience of SDSS, LSST and LHC outlines in section 4, contains some important messages for delivering SKA science and the size and nature of SKA-Data.

1. SKA surveys will produce a large range of data volumes that extend into the multi-Exabyte range. The 2020+ pricing of tape systems to manage the entirety of these surveys to Level 6, plus associated HPC costs, will exceed the anticipated SKA operational and construction budgets by factors of order 5-10. This fact alone implies multiple data release and flush cycles, and a shift of the long-term responsibility for storage to outside of SKA Observatory resources. An early realization of a similar challenge led the LHC to propose a new distributed approach to delivering science.
2. Both SDSS and LHC are examples of projects that have underestimated their data sizes and software/hardware costs. The factors involved are 4-24. There are potentially similar factors lurking in the SKA effort:

   a. The radio astronomy delivered performance of HPC verses Linpack benchmarks (factor of 14 – Figure 3)
   b. The resources needed beyond Level 6 to curate and deliver SKA science based on surveys (at least 7x Level 6)
3. SDSS is the most productive astronomical project of all time as measured by publications, citations and studentships. This was achieved by the design and development of a high performance and distributed database system that enabled collaboration between science teams. If SKA is to achieve the same success, then we need to leverage this experience and invest in a distributed and database focused, SKA survey science capability.
4. LHC realized and faced the same challenges to distribute data and do science that are now confronting the SKA. To do this, they took 10 years to design and implement an independent, distributed and complex collaborative network of research organizations (the WLCG = "the grid"). SKA will need to seriously consider this distributed concept, with all of its costs, benefits and complexity.
5. A distributed approach to doing challenging and expensive science, while complex, has been recognized by the LHC to have significant advantages in engaging and developing the community that built and operate the facility.

Is it feasible to considering distributing a large fraction of the SKA data flow and data management responsibility on the timescale of 2020+? The following figure shows the mean data flow rate in TB/s for KSP science products over the duration of a given project.

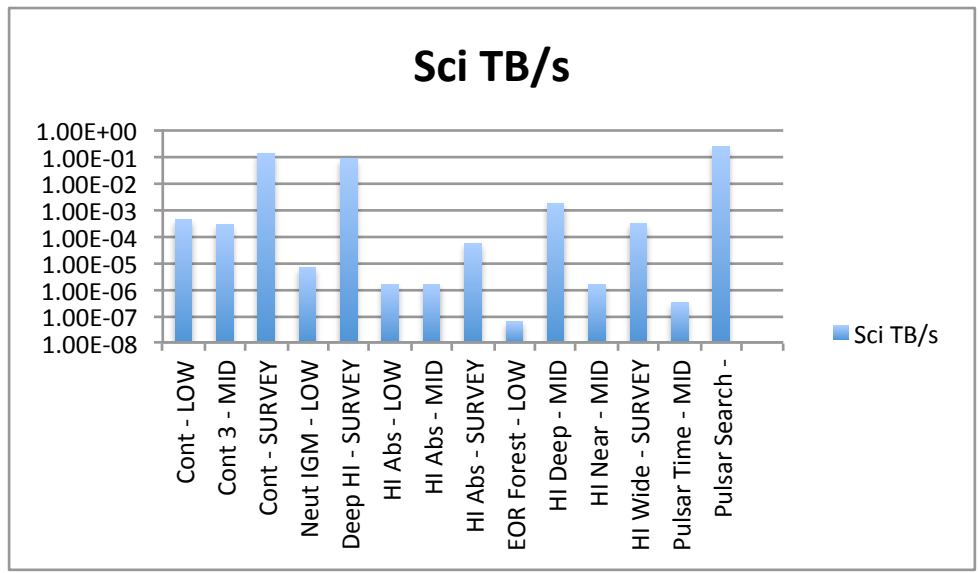

*Figure 13: The mean data flow rate in TB/sec for the KSP set averaged over each projects exposure time.*

Even for the largest surveys, the mean data flow rates are approximately 1 Tb/s. These data rates are expected to be widely available internationally and affordable on the timescale of 2020.

Will SKA need to reinvent an infrastructure like WLCG to do SKA science? Since the first planning for the WLCG in 2001, and the subsequent operational roll-out of "the grid" in the past 5-10 years, there has been a growing trend to industrialize, simplify and commercialize distributed computing through the creation of "Cloud Computing and Storage". Cloud resources have now evolved from the simple decentralize storage of flat files, to reconfigurable arrays of processing and storage capabilities of significant scale. The Amazon Web Services (AWS) set of capabilities extends from modest backup for individual desktops to entire data and processing environments for US government agencies including NSA and NASA. The SkyNet community computing projects in radio astronomy ([www.theskynet.org](www.theskynet.org)) (Vinsen 2013), based on AWS servers and storage, has already demonstrated the possibility of delivering Top500 class computing resources to astronomy applications. The cost of Cloud-based storage and processing is also evolving rapidly. Figure 14 shows the cost trends and models for AWS storage that are projected to be comparable to tape system costs per gigabyte by 2020.

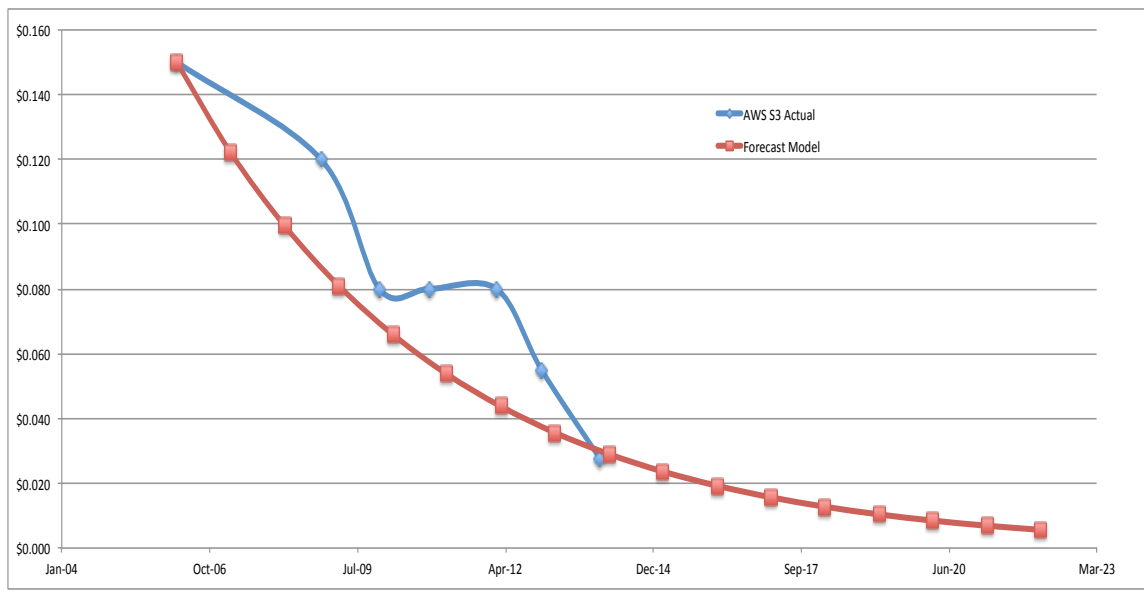

*Figure 14: The projected and actual costs of AWS storage in dollars per gigabyte (source: AmazonWebServices.com).*

The Cloud model for resources also shifts cost centres from capital expenses (initial investments in hardware) to ongoing operational costs. This strategy avoids the overheads and the direct costs associated with aging hardware, local fluctuations and growth in power costs, staffing and management investment, depreciation and the migration of content as technologies evolve. It also allows a uniform data access and analysis environment (part of AWS and others) in which physically separate research teams can simultaneously utilize the same astronomer-developed code-base, data and processing resources. While some measure of HPC and storage needs to be within the SKA Observatory and "plumbed into" the telescopes (Tier 0 in the LHC model), the distribution of analysis and survey science effort across the SKA community, may well benefit from appropriate utilization of Cloud resources, and the experience of cloud providers who are currently rolling out top 20-scale machines every few days. At ICRAR, we are currently undertaking a trial Cloud implementation of a radio astronomy survey project (the CHILES JVLA deep HI survey, PI Jacqueline van Gorkom, Columbia U) in collaboration with Amazon Web Services to better understand Cloud based work flows and performance relevant to the SKA-data design process.

## 6. Conclusions

Based on an assessment of the data flow and storage requirements of a set of SKA Key Science Projects, and their projected operational and construction costs, it would appear likely that the resources needed for the full development of SKA science data products, and the long-term storage and access to science data, will require significant community involvement and commitment. That commitment, while technically necessary and financially significant, will also strongly bind and engage the diverse SKA international community into the scientific mission and success of the SKA. The developments undertaken by the SDSS and LHC projects, to facilitate distributed research, are highly relevant to the task that the SKA community should now undertake. That task is to design and build a SKA-data system that will, through working in close coordination with the SKA Observatory, enable survey science teams to delivery on the potential of the SKA for major astronomical breakthroughs in the coming decade. We feel that this task will be significantly assisted by the roll-out of globally available Cloud technologies and services.